\documentclass[useAMS,usenatbib]{mn2e}
\usepackage{graphicx}
\usepackage{lscape}
\usepackage{color}
\usepackage{tabularx}

\usepackage{times}
\begin{document}

\title[Stellar parameters and distance to ASAS J180057-2333.8.]{The Araucaria Project. Accurate stellar parameters and distance to evolved eclipsing binary ASAS J180057-2333.8 in Sagittarius Arm. }
\author[K.Suchomska et al.]
{K. Suchomska$^1$\thanks{E-mail: ksenia@astrouw.edu.pl}, D. Graczyk$^{2,3}$, R. Smolec$^4$, G. Pietrzy{\'n}ski$^{1,3}$, W. Gieren$^{3,2}$, K. St\c{e}pie{\'n}$^1$
\newauthor  P. Konorski$^1$,  B. Pilecki$^{1,3}$, S. Villanova$^3$, I. B. Thompson$^5$, M. G{\'o}rski$^{1,2}$, P. Karczmarek$^1$, 
\newauthor P. Wielg{\'o}rski$^1$, R. I. Anderson$^{6,7}$. \\
$^1$Warsaw University Observatory, Al. Ujazdowskie 4, 00-478 Warsaw, Poland\\
$^2$Millenium Institute of Astrophysics, Av. Vicu{\~n}a Mackenna 4860, Santiago, Chile\\
$^3$Departamento de Astronom{\'i}a, Universidad de Concepci{\'o}n, Casilla 160-C, Concepci{\'o}n, Chile\\
$^4$ Nicolaus Copernicus Astronomical Centre, Bartycka 18, 00-716 Warsaw, Poland\\
$^5$ Carnegie Observatories, 813 Santa Barbara Street, Pasadena, CA 911101-1292, USA\\
$^6$ Department of Physics and Astronomy, The Johns Hopkins University, Baltimore, MD 21202, USA \\
$^7$D\'epartement d'Astronomie, Universit\'e de Gen\`eve, 51 Ch. des Maillettes, 1290 Sauverny, Switzerland}
\maketitle

\begin{abstract}
We have analyzed the double-lined eclipsing binary system ASAS J180057-2333.8 from the \emph{All Sky Automated Survey} (ASAS) catalogue . We measure absolute physical and orbital parameters for this system based on archival $V$-band and $I$-band ASAS photometry, as well as on high-resolution spectroscopic data obtained with ESO 3.6m/HARPS and CORALIE spectrographs.
The physical and orbital parameters of the system were derived with an accuracy of about 0.5 -- 3\%. The system is a very rare configuration of two bright well-detached giants of spectral types K1 and K4 and luminosity class II. The radii of the stars are $R_1$ = 52.12 $\pm$ 1.38 and $R_2$ = 67.63 $\pm$ 1.40 R$_\odot$ and their masses are $M_1$ = 4.914 $\pm$ 0.021 and $M_2$ = 4.875$\pm$ 0.021 M$_\odot$ . The exquisite accuracy of 0.5\% obtained for the masses of the components is one of the best mass determinations for giants. We derived a precise distance to the system of 2.14 $\pm$ 0.06 kpc (stat.) $\pm$ 0.05 (syst.) which places the star in the Sagittarius-Carina arm. The Galactic rotational velocity of the star is $\Theta_s=258 \pm 26$ km s$^{-1}$ assuming $\Theta_0=238$ km s$^{-1}$. A comparison with PARSEC isochrones places the system at the early phase of core helium burning with an age of  slightly larger than 100 million years. The effect of overshooting on stellar evolutionary tracks was explored using the MESA star code. 
	%The accuracy of the determination of stellar parameters is limited by quality of the light curve of this system, and with better photometry an analysis should obtain an accuracy comparable with that expected from the GAIA mission.      
\end{abstract}


\section{Introduction}

The analysis of binary star systems is a very important part of astrophysics. Calculations of their orbits allow us to directly determine the masses of their components, which also gives us a chance to estimate other physical parameters. Moreover, analysis of eclipsing binary systems can provide the absolute values for the numerous physical parameters of stars, which are essential for testing  stellar structure and evolutionary models. In particular, the SB2 eclipsing binary systems allow us to  directly determine those parameters with a requisite accuracy \citep{torres10}. From both photometric and spectroscopic data, we can measure very accurate distance-independent stellar parameters  such as stellar masses, radii, luminosities and effective temperatures. 

In this paper we present the first determination of the physical and orbital parameters of the double-lined detached eclipsing binary system from the \emph{All Sky Automated Survey} (ASAS) identified as ASAS J180057-2333.8 (hereafter ASAS1800) by \cite{poj02}. The star is also cataloged as TYC 6842-1399-1 and 2MASS J18005707-2333420, and classified as a detached binary system in the ACVS catalogue. Its V magnitude at maximum brightness is 10.19 \citep{poj02} and the amplitude of photometric variations in the $V$-band is 0.47 mag. It has a circular orbit with a period of 269 days. The system contains two evolved giants. It is located in the Galactic disc ($b=-0^{\circ}\!\!.2$) and has not been spectroscopically analyzed before. Such systems are very rarely found \citep[eg.][]{hel15} in our Galaxy. Although it is relatively young, the binary system is composed of well detached bright giant stars. Late type eclipsing binary systems are one of the best candidates for distance determinations \citep{pie13, thompson01}. The precision of our distance determination to this  eclipsing binary (total error of $\sim 4\%$) rivals those obtained from interferometric parallaxes of Galactic masers \citep[e.g.][their Table 1]{Xu12} at comparable distances.
In this paper we focus on a precise determination of physical parameters of the system, its distance and space kinematic properties and a discussion of evolutionary status. Absolute dimensions of both components are used in discussion of ASAS1800's evolutionary status. We begin with a presentation of the data collection and analysis followed by a description of our results. In the last section we present our conclusions. 

\section{OBSERVATIONS}

\subsection{Photometry}
For our analysis of ASAS1800 we used the archival $V$-band and $I$-band photometry from the ACVS \citep{poj00}. A total of 887 and 266 measurements were obtained in the $V$-band and the $I$-band, respectively, and the  data coverage for the light curve for this system is complete in both filters. The primary eclipse is total. The time span of the observations for the $V$-band is 3189 days (JD 2451949 to JD 2455138) and 1675 days for the $I$-band (JD 2452282 to JD 2453957). The magnitude in the K-band was taken from 2MASS catalogue and is K = 5.917 mag \citep{cutri03}. The observation was made at an orbital phase of   $\phi$ = 0.125, well separated from either eclipse.

\subsection{Spectroscopy}
The high-resolution spectra were collected with the ESO 3.6 telescope at La Silla Observatory, Chile equipped with the HARPS spectrograph, as well as with the Euler 1.2m telescope at La Silla, Chile, equipped with the CORALIE spectrograph. The resolution of the CORALIE spectrograph is $\sim$50,000. The  HARPS spectrograph was used in the EGGS mode at a resolution of $\sim$80,000. For our analysis we used 14 spectra in total, 12 of which were taken with the HARPS spectrograph and 2 spectra with CORALIE. 

\section{ANALYSIS AND RESULTS}
In order to derive absolute physical and orbital parameters for the system, we used the Wilson-Devinney code (WD), version 2007 \citep{wil07, wilson71, wilson79, wilson90}, equipped with the automated differential correction (DC) optimizing subroutine and Monte Carlo simulation package. The WD code allows us to simultaneously solve multiband light curves and radial velocities which is recommended as the best way to obtain a consistent model of a binary system (Wilson 2007). We also used RaVeSpAn software written by \cite{pil12} for measuring radial velocities, as well as for spectrum disentangling, used in further analysis.

\subsection{Radial Velocities}
\begin{figure}
  \centering  
  \includegraphics[width=84mm] {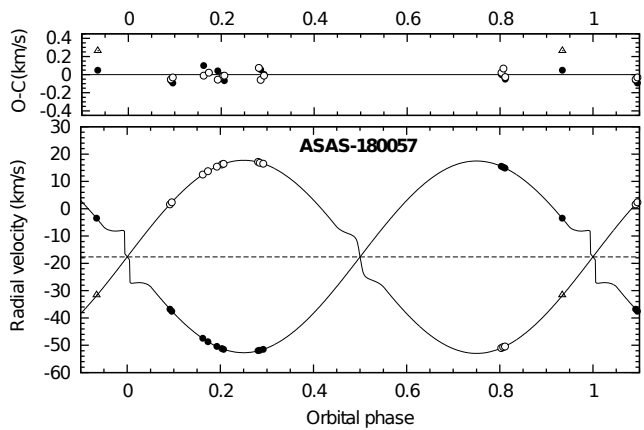}
  \caption{Radial velocity curve solution to ASAS1800 from the WD code. Filled black circles represent measurements of the primary, the empty circles measurements of the secondary. The empty triangle stands for the measurement of the secondary, observed at HJD 2456214.50228, which was not taken into account. }
  \label{fig1}
\end{figure}

The  RaVeSpAn code uses the Broadening Function formalism \citep{ruc92, ruc99} to measure radial velocities of the components of the binary. Templates were selected from the synthetic library of LTE spectra  of \citet{coelho05}. We calculated the components' radial velocities over the wavelength range 4360 to 6800 \AA, excluding atmospheric and strong hydrogen lines. The resulting radial velocity curve is presented in Fig.~\ref{fig1}, and the measured radial velocities are presented in the Tab.~\ref{tab:velocities}. The components differ in systemic velocity, and we applied a correction of $v_{sys}$= -175 m s$^{-1}$s to the radial velocities of the secondary component in order to obtain an accurate radial velocity solution. Such a shift can be caused either by a differential gravitational redshift between the stars or due to large-scale convective motions \citep[eg.][]{torres09}.

We also determined the rotational velocities of both components in order to compare them with the expected synchronous velocities. We determined rotational velocities using the RaVeSpAn code, fitting to rotationally broadened profiles. The measured broadenings are $v_{M_1}$ = 11.02$\pm$ 0.16 km s$^{-1}$ and $v_{M_2}$= 14.84 $\pm$ 0.28 km s$^{-1}$. To determine $v\sin{i}$ we also had to take into account the macroturbulence and instrumental profile contribution to our earlier measurements. In order to estimate those values we used the relation presented in \citet[][see their Equation 1]{mas08} and \cite{taked08} ($v_{mt}$ = 0.42$\zeta_{RT}$). The measured velocity is linked with rotational velocity through the relation:

\begin{equation}
\label{macro}
v_M^2 = v_{rot}^2 + (0.42\zeta_{RT})^2 + v_{ip}^2
%\log \zeta_{RT} = 3.50 \log T_{eff} + 0.25 \log (L/L_\odot) -12.67
\end{equation}
where $\zeta_{RT}$ is the radial-tangential macroturbulence, and $v_{ip}$  is the instrumental profile broadening. 
We estimated the macroturbulence to be $v_{mt1}$ = 3.20 km s$^{-1}$ and $v_{mt2}$ = 2.61 km s$^{-1}$. The instrumental profile  was estimated to be $v_{ip}$ = 2.25 km s$^{-1}$, assuming the resolution of the HARPS spectrograph in EGGS mode to be R = 80 000 and using a relation given in \citet{taked08}. With these assumptions, we estimated the rotational velocities to be $v_1\sin{i}$ = 10.31 $\pm$ 1.16 and $v_2\sin{i}$ = 14.44 $\pm$ 1.28 km s$^{-1}$.

Assuming that the rotation axes are perpendicular to the orbit and the rotation is synchronized with orbital motion, the expected equatorial, rotational velocities are $v_1$ = 9.79 km s$^{-1}$ and $v_2$ = 12.71 km s$^{-1}$. We calculated $v$ using a formula: 

 \begin{equation}
\label{v_rot}
v = \frac{2 \pi R}{P}
\end{equation}
where $R$ is the radius of a component and $P$ is the orbital period of the system.  
 The measured rotational velocities are consistent with the expected synchronous velocities within 0.5$\sigma$ and 1.4$\sigma$ for the primary and secondary components, respectively. We conclude that the components are rotating synchronously. 

\begin{table} 
\caption{Radial velocity measurements. Typical uncertainty is 40 m s$^{-1}$.}
\centering
  \begin{tabular}{p{0.22\linewidth}p{0.19\linewidth}p{0.2\linewidth}p{0.19\linewidth}}
   \hline
HJD & V$_1$ & V$_2$ &Instrument\\ 
  --2450000 & (km s$^{-1}$) & (km s$^{-1}$) &\\ \hline \hline
 5448.48918 & -36.806 & 1.770 &HARPS\\
5449.55107 & -37.554 & 2.520 &HARPS\\
5467.51220 & -47.416 & 12.664 &HARPS\\
5470.49095 & -48.694 & 13.910 &HARPS\\
5478.53899 & -51.214 & 16.369 &HARPS\\
5479.51462 & -51.498 & 16.601 &HARPS\\
5499.50211 & -51.982 & 17.304 &CORALIE\\
5500.49960 & -51.831 & 17.005 &CORALIE\\
5502.51109 & -51.492 & 16.653 &HARPS\\
6214.50228 & -3.458 & -31.415 &HARPS\\
6448.68949 & 15.524 & -50.835 &HARPS\\
6449.83022 & 15.196 & -50.472 &HARPS\\
6450.84722 & 14.856 & -50.261 &HARPS\\
6553.63729 & -50.404 & 15.572 &HARPS\\ \hline
  \end{tabular}
  \centering
  \label{tab:velocities}
\end{table}

\subsection{Spectral disentangling and atmospheric analysis}
\label{sec:dis}
The spectral disentangling was done using a method outlined by \cite{gon06}. We used the two step method described in detail in \cite{gra14} to derive properly renormalized disentangled spectra. These spectra were used for deriving the basic atmospheric parameters of effective temperature $T_{\rm eff}$, gravity $\log{g}$, microturbulance $v_t$, and metallicity [Fe/H] assuming the local thermodynamical equilibrium and using program MOOG \citep{sne73}. Details of the method are given in \cite{mar08} and the line list in \cite{vil10}. Values of these parameters  are presented in  Tab.~\ref{tab.atmo}. The derived effective temperatures of both components were used  in the WD model and to calculate interstellar reddening.

\begin{table}
\caption{Atmospheric parameters of the components.}
\begin{tabular}{p{0.22\linewidth}p{0.19\linewidth}p{0.2\linewidth}p{0.19\linewidth}}
\hline
Component & $T_{eff}$ [K] &[Fe/H] & log$g$\\ \hline
Primary & 4535 $\pm$ 70&-0.14$\pm$ 0.1 & 1.88\\
Secondary & 4240 $\pm$ 70& -0.27 $\pm$0.1& 1.93\\ \hline
\end{tabular}
\centering
\label{tab.atmo}
\end{table}
Taking into account the metallicity determination uncertainty and additional inaccuracies connected with spectra disentangling and renormalization, the difference in the metallicities of both components can be neglected (Tab.~\ref{tab.atmo}). Therefore, we can assume that the components have common metallicity within the margin of error. In our  analysis of the evolutionary status of ASAS1800 we assumed the metallicity of the system to be equal to the metallicity of the primary component - [Fe/H] = $-$ 0.14 dex (see Section~\ref{evolutionary}) .

\subsection{Interstellar extinction}
\label{sec:ext}
We estimated  reddening based on several calibrations of $T_{\rm eff}$ -  $(V - K)$ colour  \citep{ben98,alo99,hou00,ram05,gon09,cas10,mas06}, using the values of $T_{\rm eff}$ presented in Tab.~\ref{tab.atmo}. We determined E(B - V) = 0.525 $\pm$ 0.035mag. The errors on E(B-V) result from the accuracy of our effective temperature determination (see Tab.~\ref{tab.atmo}) and from the accuracy of the adopted effective temperature - colour calibrations.  color

We also determined the interstellar extinction from from the calibration of  effective temperature - $(V - I)$ colour. From the effective temperature - $(V - I)$ colour calibrations of  \cite{wor11} we estimated the intrinsic $(V - I)$ colours of each component. These colours were then compared with the observed colours of the components to estimate the E(V - I) extinction. We estimated the reddening E(V-I) for both components and then transformed it to E(B-V) using:
\begin{equation}
\label{red}
E(B-V)=\frac{E(V-I)}{1.399}
\end{equation}
The mean value of the reddening was  E(B-V)=0.609 $\pm$ 0.042 . 

Finally, we used extinction maps of \cite{schle98} with the recalibration of \cite{schla11} to estimate the reddening in the direction of ASAS1800. The total foreground reddening in this direction is E(B - V) = 26.597 mag. Since the reddening to ASAS1800 is only a fraction of this number, we have to assume a distribution of dust within the Milky Way and to know the distance to our system. The simple axisymetric model of the exponential disc gives a density of matter within the Galaxy:

\begin{equation}
\label{ro}
\rho(r, z) = \rho_0 \exp (-r/ r_d -|z|/z_d)
\end{equation} 
where $r_d$ and $z_d$ are the disc scale length and height, respectively. We adopted the following values from \cite{drim01}: sun's height in the Galactic disc $h_0$ = 0.015 kpc, $r_d$ = 3.2 kpc and $z_d$ = 0.135 kpc. The Galactic coordinates of ASAS1800 are $l = 6.37^\circ$ and $b = -0.23^\circ$. Moreover we assumed the solar distance to the Milky Way centre to be $R_0$ = 8.3 kpc \citep{gil09} and the distance to ASAS1800 to be D = 2.16 kpc (Section ~\ref{distance}). We assumed a Milky Way disc truncation at $D_{outer}$ = 20 kpc. Writing $r$ and $z$ as functions of distance $d$ from the Sun in the direction of ASAS1800 we obtain $r(d) = \sqrt{(R_0^2 + d^2 - 2R_0d\cos l)} $ and $z(d) = |h_0 + d \sin b|$. Substituting those functions into Eq. ~\ref{ro} we obtain the relation $\rho = \rho(d)$. We numerically integrated  this relation  along the line of sight twice: from 0 to D, corresponding to the reddening of the eclipsing binary, and from 0 to $D_{outer}$, corresponding to the foreground reddening. The ratio gives E(B - V)$_{ASAS1800}$ = 0.440 $\pm$ 0.057 mag. 

We adopt a final value of E(B-V) = 0.52 $\pm$ 0.07, the mean value from all our estimates. The error is a combination of both statistical and systematic error, dominated by the statistical error. 

\subsection{Modeling}
The WD code is based on Roche lobe geometry and employs a sophisticated treatment of stellar surface physics. It fits a geometric model of a detached eclipsing binary to a light curve in order to establish parameters of the system and its components.
The orbital period and the moment of primary minimum were derived from the ACVS data. We measure P = 269.363 days and $T_0$ = 2452728.5. The moment of the primary minimum (T$_0$) was later adjusted during the further analysis. The average out-of-eclipse magnitudes were established taking into account all of the observational data outside of minima. We measure V = 10.319 mag and I = 8.231 mag. Since the primary eclipse is total we were also able to  directly determine the magnitudes of the components: V$_S$=11.097 and  I$_S$=8.935 for the secondary,  and  V$_P$ = 11.061 and I$_P$ = 9.071 for the primary. We refer to the primary component as the star which is being eclipsed in the deeper, primary minimum. 

We simultaneously fitted two light curves, in the $I$-band and $V$-band, as well as the radial velocity curves. The input parameters for the DC subroutine were chosen  as described in \cite{gra12}. 
When using the Wilson-Devinney code, it is important to carefully define which parameters are adjustable in order to arrive at the best fitted model.
In our analysis we decided to adjust the orbital semi-major axis ($a$), systematic radial velocity ($\gamma$), the orbital inclination ($i$), the average surface temperature of the secondary component ($T_2$), the modified surface potential of both components ($\Omega_1$, $\Omega_2$), the mass ratio ($q = M_2/M_1$), time of the primary minimum ($T_0$), the observed orbital period ($P_{obs}$), and the relative luminosity of the primary component in the two bands ($L1_V$, $L1_I$). 

To set the effective temperature scale for each component, we ran the WD code with the initially assumed temperature of T$_1$ = 4700 K assuming a spectral type of K0 III. From the preliminary solutions obtained from the WD code we derived approximate surface gravities for the components of the binary of log $g_1$ = 1.7 and log $g_2$ = 1.4, as well as luminosity ratios in the $V$, $I$ and $K$-bands. The resulting luminosity ratios together with reddening E(B-V) (Section ~\ref{sec:ext}) were used to obtain the dereddened $(V - K)$ colour index.  The 2MASS magnitudes were converted onto the Johnson photometric system using updated transformation equations from \cite{car01}\footnote{\texttt{http://www.astro.caltech.edu/$\sim$jmc/2mass/v3/\\transformations/}} and \cite{bes88} (Tab.~\ref{tab:converted}). 
Knowing the approximate log $g_1$ = 1.7, the dereddened $(V - K)$ = 2.61 and assuming [Fe/H] = 0 dex,  we were able to estimate preliminary effective temperatures of the components based on the calibrations given by \cite{wor11}. The resulting effective temperature was set as the temperature of the primary $T_1$ = 4550 K. We used this value as a starting point for our analysis and then iterated to find the best solution for both the $V$-band and $I$-band light curves using the LC subroutine of the WD code. All free parameters were adjusted at the same time. 

The albedo and gravity brightening parameters were set to 0.5 and 0.32 respectively, which are appropriate values for those kind of stars \citep{lucy67}. To compute the limb darkening coefficients we used the logarithmic law of \cite{kling70}. Those coefficients were calculated internally by the WD code during each iteration of DC using tabulated data computed by \cite{van93}. Additionally, we calculated models using the linear and square root limb darkening law. However, that resulted in a slightly worse fit to the light curves, changing the stellar parameters of the system by less than 0.5\%. Thus, we adopted the solution obtained with fixed coefficients of the logarithmic limb darkening law \citep{pie13}. 

At the end of the fitting procedure we additionally adjusted the third light ($I_3$) to determine its impact on the solution. Formally, the solution suggested an unphysical value for $I_3$, and we therefore  set $I_3$ = 0 in our final solution. 

The solution, especially the luminosity ratio of the components, was used to renormalize the disentangled spectra. Subsequently, the atmospheric analysis was performed in order to obtain a better estimation of the temperatures of the components and their metallicities (see Section~\ref{sec:dis}). We derived effective temperatures of $T_1$ = 4535 $\pm$ 70 K and $T_2$ = 4240 $\pm$ 70 K and metallicities of [Fe/H]$_1$ = -0.14 $\pm$ 0.1 dex and [Fe/H]$_2$ = -0.27 $\pm$ 0.1 dex. We then adopted $T_1$ as the new effective temperature of the primary component and we repeated the fitting using the DC subroutine of the WD code.

\begin{figure}
  \centering  
  \includegraphics[width=84mm] {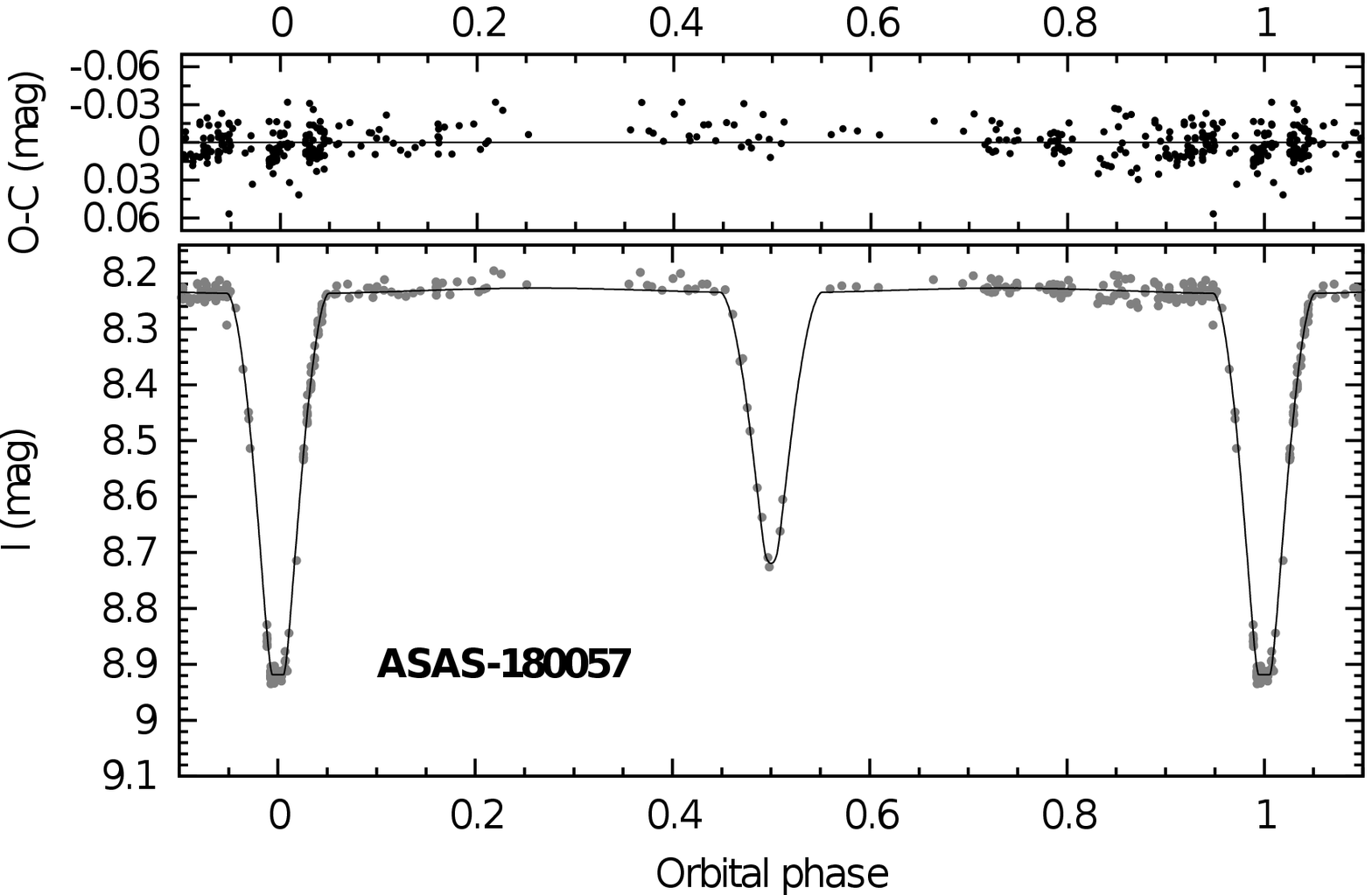}
  \caption{ The $I$-band light curve of ASAS1800 together withe the solution from the WD code.}
  \label{fig2}
\end{figure}

\begin{figure}
  \centering  
  \includegraphics[width=84mm] {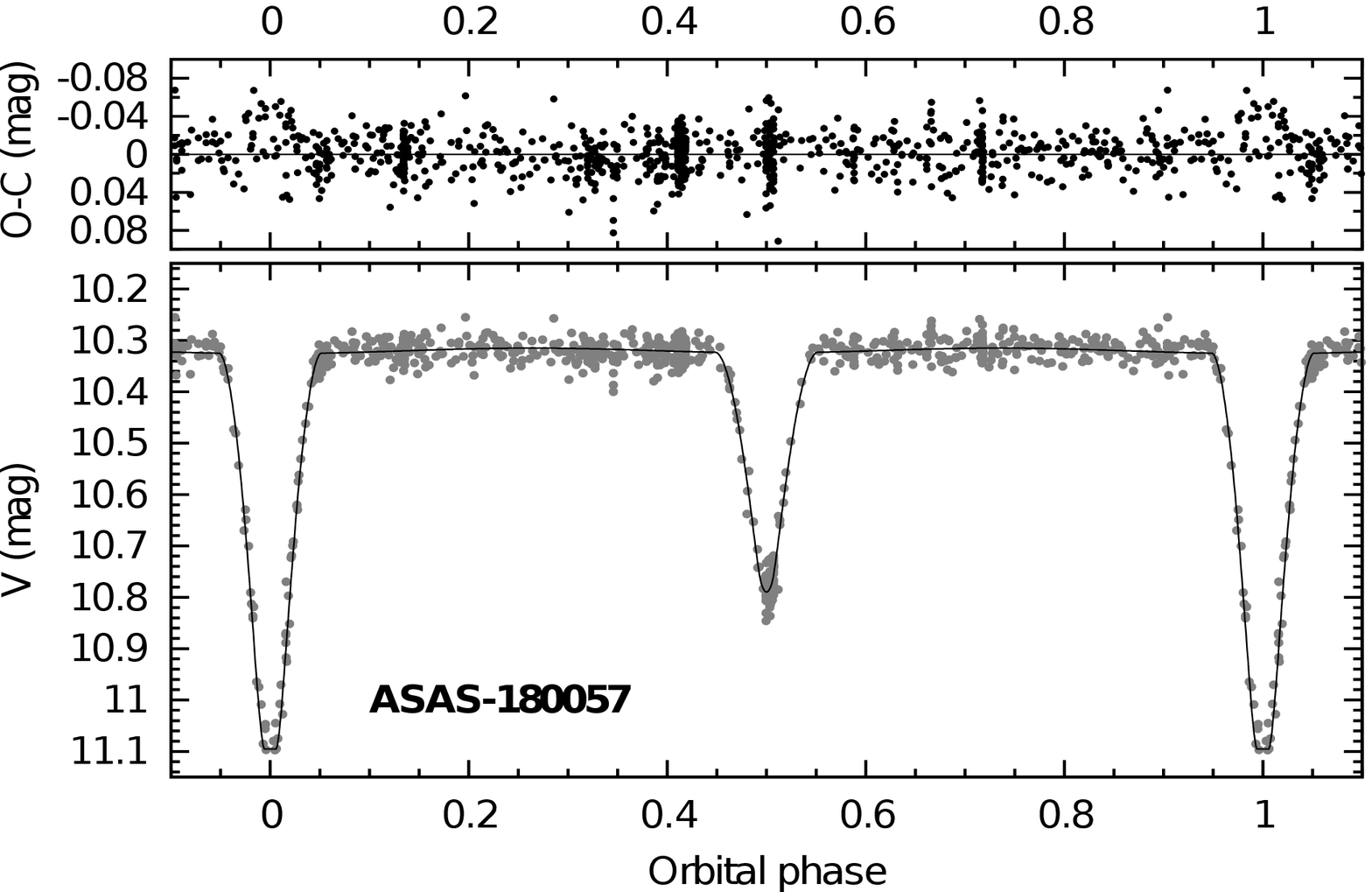}
  \caption{The $V$-band light curve of ASAS1800 together with the solution from the WD code. }
  \label{fig3}
\end{figure}

The $I$-band light curve solution obtained with the WD code is presented in Figure~\ref{fig2} and for the $V$-band in Figure~\ref{fig3} and the parameters are summarized in Table~\ref{table:results}.  

% We find our solution very satisfying, taking into account the quality of the $I$-band photometry. Nevertheless, basing on the $V$-band and $I$-band light curves that we used, it is the best solution we could obtain. 

\begin{table} 
\caption{Photometric and orbital parameters obtained with the Wilson--Devinney code.}
 \begin{tabular}{p{0.63\linewidth}p{0.26\linewidth}}
\hline
Parameter & WD result \\ \hline \hline
Orbital inclination $i$ (deg) & 88.67 $\pm$ 0.21 \\
Orbital eccentricity $e$ & 0.0 (fixed)\\
Sec. temperature $T_2$ (K) &  4211 $\pm$ 13\\
Fractional radius $r_1$ & 0.1387 $\pm$ 0.0020 \\
Fractional radius $r_2$ & 0.1800 $\pm$ 0.0012\\
$(r_1+r_2)$ & 0.3187 $\pm$ 0.0012\\
$k = r_2/r_1$ & 1.2976 $\pm$ 0.0104\\
Observed period $P_{obs}$ (day) & 269.496 $\pm$ 0.014\\
$(L2/L1)_V$ & 0.9850 $\pm$ 0.0054\\
$(L2/L1)_I$ & 1.1650 $\pm$ 0.0072 \\
$(L2/L1)_J$ & 1.3379\\
$(L2/L1)_K$ & 1.5249\\
$T_0$ (JD-2450000) & 2728.82 $\pm$ 0.06\\
Semimajor axis $a$ (R$_\odot$) & 375.72 $\pm$  0.37\\
Systemic velocity $\gamma$ (km s$^{-1}$) & -17.625 $\pm$ 0.021\\
Prim. velocity semi-amplitude $K_1$(km s$^{-1}$) & 35.11 $\pm$ 0.10\\
Sec. velocity semi-amplitude $K_2$(km s$^{-1}$) & 35.38 $\pm$ 0.10\\
Mass ratio $q$ & 0.992 $\pm$ 0.003 \\ 
RV $rms_1$  (km s$^{-1}$) &0.056\\
RV $rms_2$  (km s$^{-1}$) &0.042\\ \hline
 \end{tabular}
 \label{table:results}
\end{table}
 
 \begin{table}
  \caption{Physical Properties of the ASAS1800.}
\centering
  \begin{tabular}{p{0.3\linewidth}p{0.30\linewidth}p{0.24\linewidth}}
  \hline
  Property & The Primary & The Secondary \\ \hline \hline
  Spectral type & K1 II & K4 II\\
  $V^a$ (mag) & 11.061 & 11.098 \\
 % $I$ (mag) & 8.231 & 8.935 \\
  $V\!-\!I^a$ (mag) & 1.991 & 2.162 \\
  $V\!-\!K^a$ (mag) & 4.124 & 4.598 \\
  $J\!-\!K^a$ (mag) & 1.087 & 1.229 \\
  Radius ($R_{\odot}$) & 52.12$\pm$ 1.38 & 67.63 $\pm$ 1.40\\
  Mass ($M_{\odot}$) & 4.914 $\pm$ 0.021 & 4.875 $\pm$ 0.021\\
  log $g$ (cgs) & 1.696 $\pm$ 0.023 & 1.466 $\pm$ 0.018  \\
  $T_{\rm eff}$ (K) & 4535$^b$ $\pm$ 80 & 4211$^c$ $\pm$ 80 \\
  $v$ sin $i$ (km s$^{-1}$) & 10.31 $\pm$ 1.16 & 14.44 $\pm$ 1.28  \\
  Luminosity ($L_{\odot}$) & 1031 $\pm$ 91& 1290 $\pm$ 111\\
  $M_{\rm bol}$ (mag) & -2.80 & -3.05\\
  $M_{\rm v}$ (mag) & -2.33 & -2.32 \\
  $[$Fe/H$]^b$ & -0.14 $\pm$ 0.1& -0.27 $\pm$ 0.1\\ \hline
  $E(B\!-\!V)$ & 0.525 $\pm$ 0.07 \\
  Distance (pc) & 2142.5 $\pm$ 63.5 (stat.) & $\pm$ 53.3 (syst.) \\ 
 \hline
 $^{a - observed}$ 
 $^{b- atmospheric\: analysis}$
 $^{c -WD\: solution}$ 
  \end{tabular}
  \centering
  \label{tab:results_final}
\end{table}

 \begin{table*}
 \centering
  \caption{Error budget of the distance moduli of the ASAS1800.}

 \begin{tabular}{@{}ccccccclcc@{}}
  \hline
    Type of error & $(m - M)$ & $\sigma$A  & $\sigma$(MonteCarlo) & $\sigma$diBenedetto & $\sigma E(B-V)$  & $\sigma V$ & $\sigma K$  & $(L_2/L_1)_K$ &Combined Error  \\
    &(mag)&(mag)&(mag)&(mag)&(mag)&(mag)&(mag)&(mag) &(mag) \\ \hline \hline
    \textbf{Statistical}&11.655&0.003&0.049&--&0.024$^1$&0.0241&0.019& -&\textbf{0.063}\\
   \textbf{Systematic} &11.655&--&--&0.043&--&\multicolumn{2}{c}{0.03}&0.01&\textbf{0.053} \\
  \hline
  \multicolumn{3}{c}1 - combination of statistical and systematic error
  \end{tabular}
  \centering
  \label{tab:error}
\end{table*}

\subsection{Absolute Dimensions}
%The observed orbital period of the system $P_{obs}$ and the true orbital period $P$ are linked through a relation:

%\begin{equation}
%P_{obs}= P (1+\frac{\gamma}{c})
%\end{equation}
%where $c$ is the velocity of the light and $\gamma$ is the systemic velocity. 

Table~\ref{tab:results_final} gives astrophysical data about the two components. The physical radii of the stars result from the relation: $R = r \cdot a$, where $r$ is the fractional radius listed in Table~\ref{table:results}.  The masses are derived from the equations:
\begin{equation}
M_1[M\odot] = 1.32068 \cdot 10^{-2} \frac{1}{1+q}\frac{a^3[R\odot]}{P^2[d]}
\end{equation}
\begin{equation}
M_2 [M\odot] = M_1 \cdot q
\end{equation}
where $a$ is the semi major axis, $q$ is the mass ratio and $P$ - real period.  
The observed individual magnitudes in both $V$ and $I$-band and $V-I$ colour relation were derived directly from the flat bottom minimum of the secondary component.
We used bolometric corrections from \cite{alo99} to convert $V$-band magnitudes into bolometric magnitudes. 
\begin{table*}
 \centering
  \caption{Out of eclipse magnitudes of ASAS1800.}
\begin{tabular}{@{}cccccc@{}}
  \hline
    & V (mag) & I (mag)& J (mag) & K (mag)  \\ \hline \hline
  ASAS1800&10.319 $\pm$ 0.024  &8.231 $\pm$ 0.015 &  7.178 $\pm$ 0.030$^1$& 5.942 $\pm$ 0.030$^1$ \\
   Reference & this work (ASAS)  &this work (ASAS)& \cite{cutri03} (2MASS) & \cite{cutri03} (2MASS)  \\
  \hline
  \multicolumn{4}{r}1 - transformed to Johnson photometric system using equations from \cite{bes88} and \cite{car01}
  \end{tabular}
  \centering
  \label{tab:converted}
\end{table*}

\subsection{Evolutionary status of ASAS1800}
\label{evolutionary}

\begin{figure*}[p]
\centering
\includegraphics[width=12.8cm]{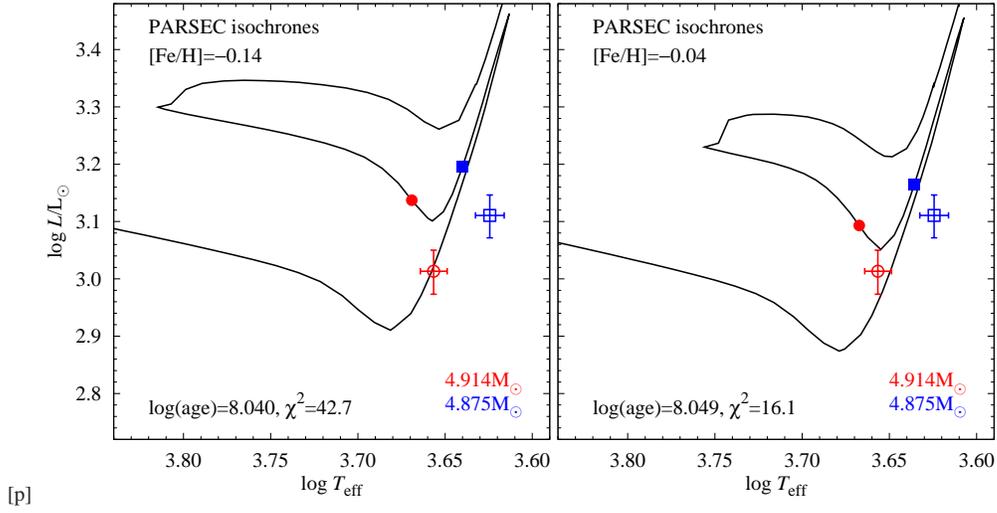}
\caption{PARSEC isochrones for two metallicities, ${\rm [Fe/H]}=-0.14$ (left panel) and ${\rm [Fe/H]}=-0.04$ (right panel). Location of primary and secondary components is marked with circles and squares, respectively. Filled symbols refer to the fitted isochrones and the empty ones to our measurements. }
\label{fig.parsec}
\end{figure*}

In this Section we compare the physical parameters of ASAS1800 (Tab.~\ref{tab:results_final}) with results of stellar evolution calculations. We assume that the components of the system have common metallicity, equal to the metallicity of the primary ($-0.14$), or possibly $1\sigma$ higher ($-0.04$). As we show below, lower metallicities lead to serious disagreement between the models and observations. In this initial study we also assume that the masses of ASAS1800, as determined in Tab.~\ref{tab:results_final}, are exact. In Fig.~\ref{fig.parsec} we plot the PARSEC isochrones \citep{bres12} for the two metallicities considered. The isochrone  was selected to minimize the $\chi^2$ function including luminosities, effective temperatures and radii of the two components. Model values were calculated at the mass points corresponding to the masses of the ASAS1800 components (filled circle and filled square for primary/secondary in Fig.~\ref{fig.parsec}).

The comparison with the PARSEC isochrones places the system at the early phase of core helium burning. The agreement with isochrones is not satisfactory, however. It is better for higher metallicity, but still the primary component is under-luminous and the secondary component is too cool and/or under-luminous. We note, however, that the PARSEC isochrones are available for only one fixed set of overshooting parameters, which strongly affect the evolutionary calculations. Below, we express the extent of overshooting as a fraction of the local pressure scale height, $\beta\times H_p$. Both the overshooting from the hydrogen-burning core during main sequence evolution ($\beta_{\rm H}$) and overshooting from the convective envelope ($\beta_{\rm env}$) affect the extent and the luminosity of the helium burning loops \citep{alongi91}. In PARSEC models, in the  mass range considered, these are fixed at $\beta_{\rm H}=0.5$ and $\beta_{\rm env}=0.7$ {\it across} the border of the convective zone determined with the Schwarzschild criterion. In the calculations described below, the extent of overshooting is measured {\it above/below} the border of the convective region, which is a more common approach. The resulting overshoot parameters roughly correspond to half of those adopted in PARSEC \cite[see discussion in section 2.6 in][]{bres12}.

\begin{figure*}[p]
\centering
\includegraphics[width=12.2cm]{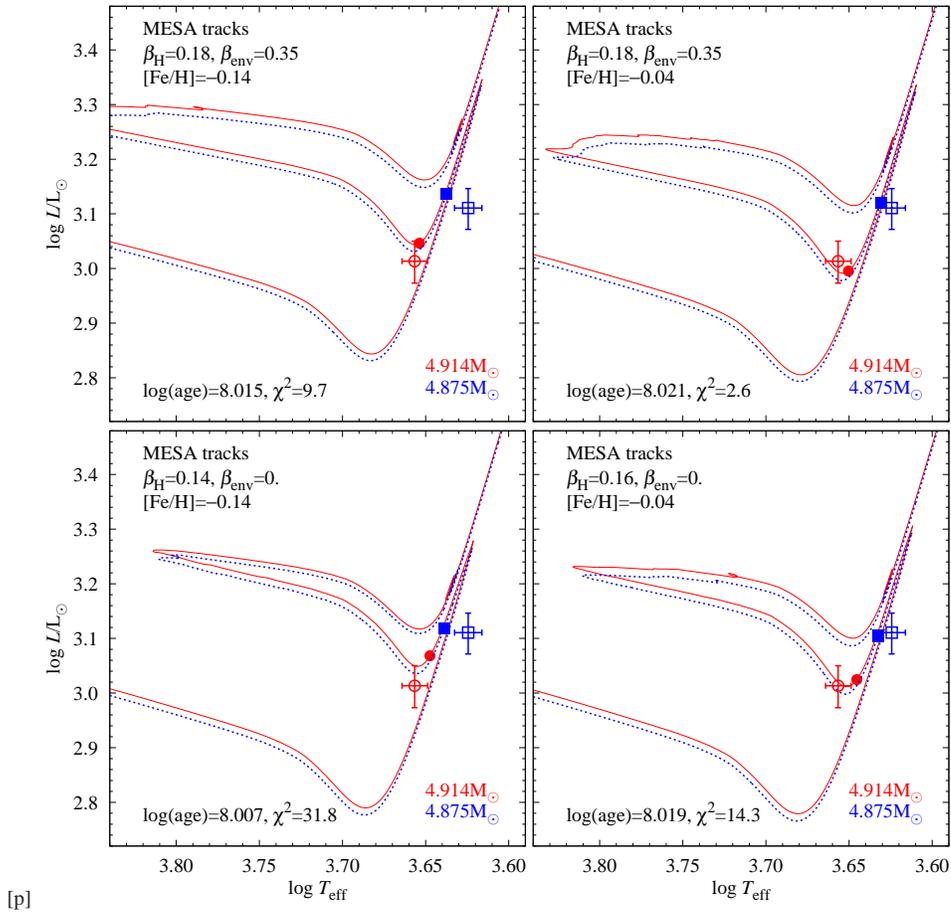}
\caption{MESA tracks computed for the primary ($M=4.914M_\odot$, red, solid line) and secondary ($M=4.875M_\odot$, blue, dashed line) components of ASAS1800, assuming the same values of overshooting from the hydrogen burning core for the primary and the secondary. Models that match the observational constraints best (at the same age) are marked with filled circles/squares for the primary/secondary. Models in the top two panels include convective envelope overshoot (neglected in the bottom panels). Metallicity is equal to ${\rm [Fe/H]}=-0.14$ (left panels) or ${\rm [Fe/H]}=-0.04$ (right panels).}
\label{fig.mesa1}
\end{figure*}

\begin{figure*}[p]
\centering
\includegraphics[width=12.2cm]{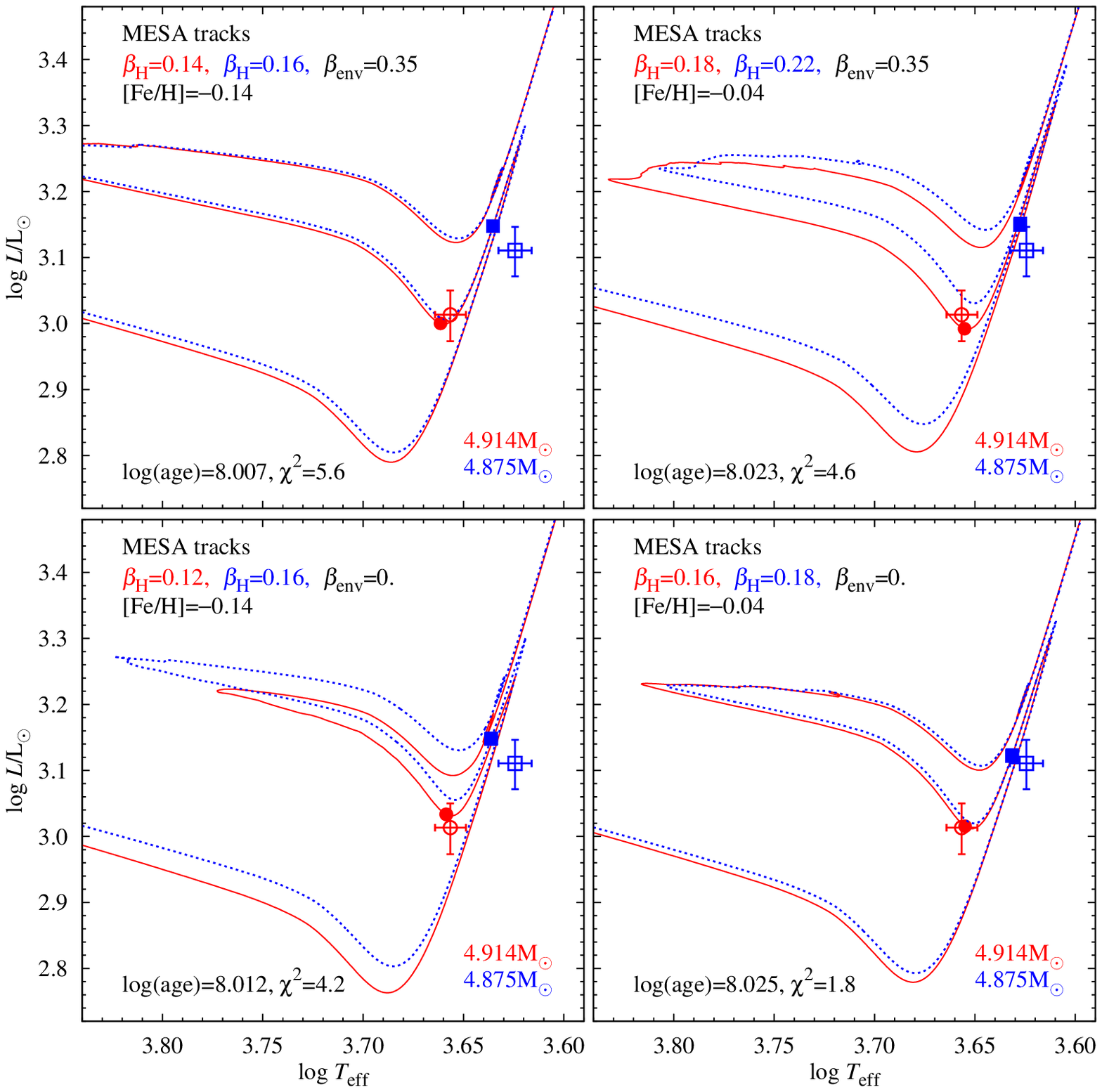}
\caption{The same as Fig.~\ref{fig.mesa1} but for models assuming different values of overshooting from the hydrogen burning core for the primary and secondary.}
\label{fig.mesa2}
\end{figure*}

To explore the effect of the overshooting on stellar evolutionary tracks we used \textsf{MESA star} -- a publicly available stellar evolution code \cite[release 6208;][]{pax11,pax13}. Details of the code setup will be described elsewhere (Smolec et al. in prep.), here we summarize the most important settings. We use OPAL opacities and adopt the \cite{asp09} solar distribution of heavy elements. Convection is modelled with the mixing length formalism \citep{boh58} with the mixing length parameter resulting from calibration of the solar model ($\alpha=1.78$). The convective boundaries are determined with the Schwarzschild criterion. We account for the overshooting above the border of hydrogen burning core, above the border of helium burning core ($\beta_{\rm He}=0.01$, fixed), and below the border of the convective envelope. We neglect rotation, element diffusion (except in solar calibration), and mass loss. For each component of ASAS1800 we fix the mass (Tab.~\ref{tab:results_final}) and compute the evolution from the pre-main sequence until the late AGB phase. Our small model grid consists of two metallicity values, $-0.14$ and $-0.04$, eight values of $\beta_{\rm H}$, $\beta_{\rm H}\in[0.1,\ 0.12,\ 0.14,\ 0.16,\ 0.18,\ 0.20,\ 0.22,\ 0.24]$, and two values of $\beta_{\rm env}$, $\beta_{\rm env}\in[0.,\ 0.35]$. Along each pair of tracks (for the two components) we determined the models that minimize $\chi^2$ function including effective temperatures, luminosities and radii of the two components, at the same age. We first assume that the two components experienced the same extent of mixing at the edge of hydrogen burning core during their evolution, and hence have the same value of $\beta_{\rm H}$, and then allow for a difference in $\beta_{\rm H}$. In Figs.~\ref{fig.mesa1} and \ref{fig.mesa2} we show our best solutions for the described two assumptions. In the four panels of these Figures we show the models with (top) and without (bottom) overshooting from the convective envelope ($\beta_{\rm env}=0.35$ or $\beta_{\rm env}=0.$) and adopting lower (left) and higher (right) metallicity values (${\rm [Fe/H]}=-0.14$ or ${\rm [Fe/H]}=-0.04$). 

We first analyze the models assuming the same values of $\beta_{\rm H}$ for the primary and secondary, Fig.~\ref{fig.mesa1}. For higher metallicity (right panels) the helium burning loops become less luminous and the tracks shift toward lower effective temperatures. Consequently, the higher metallicity (${\rm [Fe/H]}=-0.04$), together with the smaller extent of overshooting from the hydrogen-burning core, mitigates the problems of an under-luminous primary and a too cool/under-luminous secondary, noted in the analysis of the PARSEC isochrones. The inclusion of the envelope overshoot in the models has two apparent effects on the tracks. It increases the vertical extent of the loops and decreases their overall luminosity. Hence, in the models including envelope overshooting, a larger extent of overshooting at the hydrogen-burning core is possible. We note that the matter that overshoots the convective envelope boundary faces a stabilizing stratification gradient. Whether the significant mixing is possible in such case is a subject of debate \citep{bres12,pietri04}. Modelling of  evolved binary systems offers the best opportunity to test  mixing scenarios at the bottom of the convective envelope. 

The best models displayed in Fig.~\ref{fig.mesa1} are those with overshooting from the convective envelope ($\beta_{\rm env}=0.35$) and have $\beta_{\rm H}=0.18$ (Fig.~\ref{fig.mesa1}, top). The solutions assuming $\beta_{\rm H}=0.16$ are only slightly worse. When envelope overshoot is neglected, lower values of $\beta_{\rm H}$ ($\approx 0.14-0.16$) are necessary. Clearly, the best models have higher metallicity. The inferred system's age, given in Fig.~\ref{fig.mesa1}, is very similar for all models, log(age) is within a narrow range from 8.007 to 8.021. As expected, the age is slightly larger for higher metallicity models (see e.g. \cite{salaris06}). Also, larger the extent of overshooting from the hydrogen burning core, longer the main sequence evolution.

We note that for the best models in Fig.~\ref{fig.mesa1} we nearly match the luminosity of the secondary, but that the primary component is still {\it before} the observed position. Hence, we can get a much better agreement between the models and observations assuming different values of overshooting from the hydrogen burning core during the main sequence evolution. A lower value of $\beta_{\rm H}$ for the secondary slows down its evolution (extends the main sequence phase) and allows a much better match of the system with observations at the helium burning phase -- Fig.~\ref{fig.mesa2}. An overshooting parameter higher by $0.02-0.04$ for the secondary allows  very good fits. The best are obtained for higher metallicity models (as in Fig.~\ref{fig.mesa1}). This time the models that neglect the overshooting from the convective envelope seem slightly better. The inferred ages are very similar to those reported in Fig.~\ref{fig.mesa1}. Our model that matches observations best (Fig.~\ref{fig.mesa2}, bottom right), assumes $\beta_{\rm H}=0.16$ for the primary and  $\beta_{\rm H}=0.18$ for the secondary and places the system at the early helium burning phase. 

The different extent of overshooting adopted for the components of the eclipsing binary system, with nearly equal masses and metallicities of the components, might appear unjustified. We note, however, that an overshooting parameter expresses our ignorance about all kinds of mixing processes that may occur at the edge of the convective core. In particular, rotation leads to additional mixing, the extent and efficiency of which depend on the rotation rate. We see no reason to assume that the initial rotation rate was the same for two stars.

We conclude that ASAS1800 is at early phases of core helium burning. The age of the system is slightly larger than 100 million years. The models favour a metallicity that is close to solar in value.

\subsection{Tidal evolution of the system}
\label{tidal}

The observations indicate that the orbit of ASAS1800 is circular and
rotation of both components is fully synchronized with the orbital
period, which means that the memory of the initial values of
  these parameters is at present entirely forgotten. To check, if
  this fact agrees with the predictions of the tidal evolution theory
of binary stars, we assume that the binary evolves as an isolated
system with conserved total mass and angular momentum.

The tidal circularization and synchronization time scales are
(\citealt{zahn89}, \citealt{mei06})

\begin{equation}
\tau_{\rm{circ}} =
\frac{t_f}{21\lambda_{\rm{circ}}q(1+q)}\left(\frac{a}{R}\right)^8\,,
\end{equation}

\begin{equation}
\tau_{\rm{sync}} =
\frac{It_f}{6\lambda_{\rm{sync}}q^2MR^2}\left(\frac{a}{R}\right)^6\,,
\end{equation}

where $M$, $R$ and $I$ are respectively mass, radius and moment of
inertia of the tidally distorted component, $q$ is the mass ratio
(with $M$ in the denominator), $t_f$ is the viscous dissipation time,
$\lambda_{\rm{circ}}$ and $\lambda_{\rm{sync}}$ are constants
  that depend on the internal structure of a star. Both time
  scales depend on a high power of the ratio of star separation to
stellar radius. This ratio was of the order of $10^2$ when the binary
was on the main sequence (MS), resulting in both time scales much
longer than the MS lifetime of the components. Thus, the mutual tidal
interaction at that evolutionary stage can be neglected. Substantially
stronger interaction is expected when both components moved to the red
giant region. Presently, they are both past the red giant tip, burning
helium in their cores. We note that the ratio of both time scales for
our binary is $\tau_{\rm{circ}}/\tau_{\rm{sync}} \approx 25-30$ for $q
\approx 1, \lambda_{\rm{circ}} \approx \lambda_{\rm{sync}}$ and $I
\approx 0.15MR^2$ \citep{rut88}. That means that by the time the orbit
becomes circularized, the components already rotate synchronously.

To follow the eccentricity change, detailed calculations of the
circularization rate for an evolving binary are needed. Such
calculations have been performed by several authors for different
kinds of systems and upper limits for periods of fully circularized
binaries were obtained. We use the data from the paper by \citet{ver95}
who calculated the limiting period values for binaries composed of
giants. For giant masses corresponding to ASAS1800 the limiting period
is equal to 616 d (see their Table~1). Because the period of ASAS1800,
equal to 269 d, is significantly shorter than that value, we can conclude
that the zero eccentricity of its orbit is to be expected.

This conclusion can additionally be verified by a direct estimate of
the absolute value of $\tau_{\rm{circ}}$. We use to this purpose an
approximation given by \citet{ver95}

\begin{equation}
\frac{1}{\tau_{\rm{circ}}} \equiv
\left|\frac{\rm{d}\ln{e}}{\rm{d}t}\right| \approx
3.4f\left(\frac{T_e}{4500}\right)^{4/3}M_{env}^{2/3}M^{-1}\left(\frac{R}{a}\right)^8
\quad \rm{yr}^{-1}\,.
\end{equation}

We assumed $q \approx 1$. If we additionally assume that the
characteristic value of $R/a \approx 0.2$ over the giant phase, the
convection envelope mass $M_{env} \approx M$, which is a good
approximation for the first ascend giants, $T_e \approx (T_1+T_2)/2$
and $f \approx 1$ \citep{zahn89}, we obtain $\tau_{\rm{circ}} \approx
10^5$ years if tides on both components are taken into account. This
is 1-1.5 order of magnitude shorter than the lifetime of each
component of ASAS1800 in the red giant phase so the orbit was
  efficiently circularized soon after the stars reached the red
  giant branch. Because the synchronization time scale is still much
  shorter, as is shown above, the rotation of both components was
  synchronized even faster.

\subsection{Distance to the system}
\label{distance}
To derive the distance we followed prescriptions given in \cite{gra12, gra14}. We used $V$-band surface brightness (SF) - $(V\!-\!K)$ color calibration measured by \cite{ben05} for Galactic late type giant stars. The angular diameter of a star can be estimated using the formula:
\begin{equation}
\phi [{\rm mas}] = 10^{0.2(S - m_0)}
\end{equation} 
where $S$ is the surface brightness in a given band and $m_0$ is the dereddened magnitude of a star in this band. We can then directly derive the distance to the star by scaling the angular diameter:
\begin{equation}
d [pc] = 9.2984  \cdot \frac{R [R_{\odot}]}{\phi [{\rm mas}]}
\end{equation}

 The resulting distance to ASAS1800 is d = 2142.5 $\pm$ 63.5 (stat.) $\pm$ 53.3 (syst.) pc (Tab.~\ref{tab:results_final}). The main contribution to the statistical uncertainty are random errors connected with light curve modelling by the WD code (connected with a relatively large dispersion of ASAS light curves) and infrared photometry errors. Thus there is a significant room for improvement on the derived distance once high accuracy photometry will be available.. The main contribution to the systematic error comes from the SF calibration itself. The total error budget is presented in Tab~\ref{tab:error}.

\subsection{Space position and velocity}
The proper motion of the star is $\mu_{\alpha}\cos{\delta}=+0.9\pm1.75$ mas yr$^{-1}$ and $\mu_{\delta}=0.67\pm1.71$ mas yr$^{-1}$ and was derived as weighted mean from three catalogues PPMXL catalog  \citep{Roe10}, UCAC4 catalog \citep{zach13} and SPM4.0 catalog \citep{girard11}. This proper motion in  Galactic coordinates is $(\mu_l\cos{b},\mu_b)=(1.0 \pm 2.3, -0.5 \pm 0.7)$ mas yr$^{-1}$ using the prescription given by \cite{pol13}. The calculated distance corresponds to a transverse velocity in Galactic coordinates of $(10\pm24,-5\pm7)$ km s$^{-1}$. To calculate Galactic space velocity components we used equations given in \cite{jon87} and we obtained $(u,v,w)=(-19\pm3,8\pm24,-4\pm7)$ km s$^{-1}$. This velocity is not corrected for solar motion with respect to the Local Standard of Rest (LSR). Taking into account the peculiar solar motion $(U_\odot,V_\odot,W_\odot) = (11.1\pm1.0,12.2\pm2.0,7.3\pm0.5)$ km s$^{-1}$ \citep{sch10} and the circular speed of LSR in the Galaxy $V_c=238\pm9$ km s$^{-1}$ from \cite{sch12}  we obtain Galacto-centric velocity components of ASAS1800 in the position of the sun $(U_1,V_1,W_1)=(-8\pm4,258\pm26,3\pm7)$ km s$^{-1}$, where the errors are dominated by proper motion uncertainties. The Galactic space position of the star with respect to the sun is $(X,Y,Z)=(2.12,0.26,-0.05)$ kpc. The Galacto-centric distance of ASAS1800 is $6.16\pm0.40$ kpc \citep[assuming a distance to the Galactic center $R_0=8.28 \pm 0.38$ kpc from][]{gil09} and the Galacto-centric longitude is $\beta = 2^{\circ}\!\!.5\pm0^{\circ}\!\!.3$, placing the star in the Saggitarius-Carina arm \citep[e.g.][their Figure 3]{sak12}. 

\section{SUMMARY AND CONCLUSIONS}
 We have obtained stellar parameters for the eclipsing binary ASAS J180057-2333.8. We measure a  distance to the system of 2.14 $\pm$ 0.06 (stat.) $\pm$ 0.05 (syst.) kpc. The accuracy of the distance determination is 4 \%, is slightly less accurate than distances obtained with the same method to LMC/SMC binaries. This is due to much higher interstellar extinction, somewhat lower quality of the photometric light curve and infrared magnitudes transformations between photometric systems, leading to larger errors in the absolute dimensions and final distance. With better photometry and with an improved surface brightness - colour relation it should be possible to measure 1.5 -- 2 \% distances to such individual systems. In a recent series of papers  \citep{pie09,gra12,gra14} we have shown that precision of 3\% is already routinely attainable for carefully selected late type eclipsing binaries. As such they are a useful tool to probe the structure and the kinematics of the Galaxy. This technique is also an important and, moreover, independent way of testing the future distance and parallax determinations which will be made by the GAIA mission.
 
Our results also demonstrate the strength of using observations of  well detached eclipsing binary systems in the testing of stellar evolution theory. Several such systems, with well determined physical parameters, are known \cite[eg.][]{pie13}. Evolutionary calculations of evolved stars are sensitive to many parameters however, and definite conclusions require a thorough study, which is ongoing (Smolec et al. in prep.).

\section*{Acknowledgements}
We would like to thank the staff of the ESO La Silla observatory for their support during the observations. We also gratefully acknowledge financial support for this work from the Polish National Science Centre grants OPUS DEC-2013/09/B/ST9/01551 and DEC-2011/03/B/ST9/02573 and the TEAM subsidy from the Foundation for Polish Science (FNP). In this work we used SIMBAD database. W.G., G.P. and D.G gratefully acknowledge financial support for this work from the BASAL Centro de Astrofisica y Tecnologias Afines (CATA) PFB-06/2007, and from the Millenium Institute of Astrophysics (MAS) of the Iniciativa Cientifica Milenio del Ministerio de Economia, Fomento y Turismo de Chile, project IC120009. K.S. acknowledges the financial support from the National Science Centre under the grant DEC-2011/03/B/ST9/03299. RIA acknowledges funding from the Swiss National Science Foundation. RIA, WG and GP acknowledge the support of the Munich Institute for Astro- and Particle Physics (MIAPP) of the DFG cluster of excellence "Origin and Structure of the Universe".

\end{document}